\documentclass[sn-mathphys]{sn-jnl}
\usepackage[superscript,biblabel]{cite}

\begin{document}

\title[Black Holes Up Close]{Black Holes Up Close}


\author*[1,2]{\fnm{Ramesh} \sur{Narayan}}\email{rnarayan@cfa.harvard.edu}

\author[3]{\fnm{Eliot} \sur{Quataert}}\email{quataert@princeton.edu}

\affil*[1]{\orgdiv{Center for Astrophysics}, \orgname{Harvard \& Smithsonian},  \orgaddress{\street{60 Garden Street}, \city{Cambridge}, \postcode{MA 02138}, \country{USA}}}

\affil[2]{\orgdiv{Black Hole Initiative}, \orgname{Harvard University}, \orgaddress{\street{20 Garden Street}, \city{Cambridge}, \postcode{MA 02138}, \country{USA}}}

\affil[3]{\orgdiv{Department of Astrophysical Sciences}, \orgname{Princeton University},  \city{Princeton}, \postcode{NJ 08544},  \country{USA}}


\abstract{Recent developments have ushered in a new era in the field of black hole astrophysics, providing our first direct view of the remarkable environment near black hole event horizons.  These observations   have enabled astronomers to confirm long-standing ideas on the physics of gas flowing into black holes with temperatures that are {hundreds} of times greater than at the center of the Sun. At the same time, the observations have conclusively shown that light rays near a black hole experience large deflections which cause a dark shadow in the center of the image, an effect predicted by Einstein's theory of General Relativity. With further investment, this field is poised to deliver decades of advances in our understanding of gravity and black holes through new and stringent tests of General Relativity, as well as new insights into the role of black holes as the central engines powering a wide range of astronomical phenomena.}

\maketitle

\section*{}\label{sec1}

{Einstein's theory of General Relativity (GR) radically altered our understanding of the nature of space and time.   In GR, space and time are dynamical quantities, leading to the existence of gravitational waves and the expanding Universe.  But perhaps most remarkably, GR  predicts a fundamentally new type of object, the black hole (BH).   Unlike normal stars and planets which have surfaces, BHs in GR are defined by the presence of an event horizon, a region within which gravity is so strong that nothing can escape {\cite{Schwarzschild1916,Kerr1963,Misner1973}}. (There were Newtonian physics-based predictions of BH-like objects in the late 1700s by Michell {\cite{Michell1784}} and Laplace {\cite{Laplace1799}}, but their prescient ideas were too far ahead of the times and were not subsequently developed.)

Although initially BHs (much like gravitational waves) were regarded as mathematical curiosities or even artifacts, starting in the 1960s BHs became a focus of research in both physics and astronomy {\cite{Thorne1994,BR2010}}. 

In physics, BHs have played a starring role in the theoretical quest for a quantum theory of gravity, in part because GR predicts its own failure in the interior of BHs:  all of the matter out of which the BH is made collapses to a singularity of infinite density where the equations of GR break down. ({Roger Penrose received a share of the 2020 Nobel Prize in Physics for his seminal paper {\cite{Penrose1965}} showing that such singularities generically form during gravitational collapse.)}  {There are also major theoretical puzzles reconciling quantum mechanics and GR.  The most traction in elucidating this tension has occurred in the context of understanding Hawking's famous prediction {\cite{Hawking1975}} that in quantum mechanics BHs should actually produce a small amount of radiation (which, we note, is completely unobservable for any BH that has been discovered!).}

{In astronomy, a sixty year quest has led to the amazing realization that BHs are not only real, but common.  There are tens of millions of stellar-mass BHs per galaxy:  a few of these happen to shine as bright X-ray sources, and gravitational wave measurements have detected dozens of mergers of stellar mass BHs.   In addition, right at the center of each galaxy, there is {usually a} supermassive BH with a mass $\sim 10^5-10^{10} M_\odot$ (where $M_\odot \simeq 1.99 \times 10^{33}$\,g is the mass of the Sun)}.

{Ever since the earliest discovery of BHs as quasars (enigmatic radio and optical sources at the centers of galaxies {\cite{Schmidt1963,Salpeter1964})}, the quest was on to directly see the environment near the event horizon.   In the most opportune cases, theorists predicted that such images would be dramatic {\cite{Luminet1979, Falcke2000}}:  a deficit of radiation from near the event horizon due to the strong bending of light by the BH's gravity, together with  asymmetries in the image because half the radiating matter near the BH would be moving towards us (and thus appear brighter due to the special relativistic Doppler shift) and half would be moving away from us (and thus appear fainter).  As shown in Figure \ref{fig:EHT_images}, those predictions have now been  confirmed {\cite{EHT_M87*, EHT_SgrA*}}. {In particular, seeing the dark ``shadow" at the center of the image and the ring of emission around it is a remarkable achievement which shows that our GR-inspired understanding of BHs is fundamentally correct.} 

\begin{figure}[!ht]%
\centering

\includegraphics[width=0.99\textwidth]{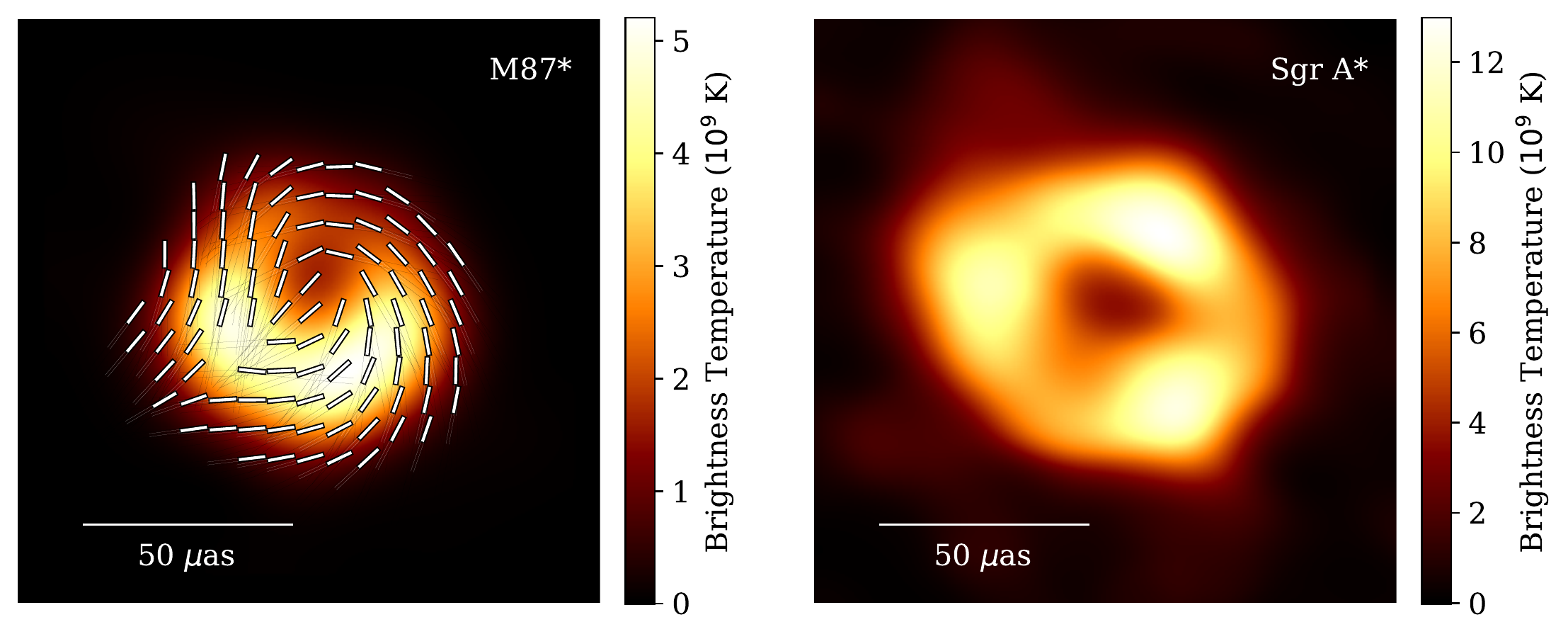}

\caption{
 {{\bf Event Horizon Telescope  images of two supermassive BHs.} {\bf Left}, Image of M87* ($\mu{\rm as} =$ microarcsecond), where color scale corresponds to the ``brightness temperature" of the observed radiation, and tick marks indicate the orientation of linear polarization. {\bf Right}, Image of Sgr A*. Both images show a dark central region, called the ``shadow'' of the black hole, and an asymmetric ring of emission around it.  Figure credit: Daniel Palumbo, adapted from published EHT results {\cite{EHT_M87IV, EHT_M87VII,EHT_SgrAIII}}.}}
\label{fig:EHT_images}
\end{figure}

Making these images was, however, extraordinarily challenging:   it required a telescope with a thousand times better {angular resolution} than the James Webb Space Telescope and was akin to taking a picture of a tennis ball 
on the Moon using a camera on Earth.  Many decades of theoretical, telescope, and software development led to these remarkable results.   We now provide additional context for understanding event-horizon scale physics, what observations have been made to date, and what the future may hold.}

\section*{Black Hole Accretion and Gravity}

{This perspective will focus primarily on two of the most well-studied supermassive BHs:  the $4 \times 10^6 M_\odot$ BH, called Sagittarius A* (Sgr A*), at the center of our own Galaxy {\cite{Eckart1996, Ghez1998, Ghez2000, Schodel2002,Ghez2005}}, and the $\simeq 6 \times 10^9 M_\odot$ BH, called M87* {\cite{Gebhardt2011}}, at the center of the galaxy M87. The case for Sgr A* being a BH, {based on measurements of the orbits of surrounding stars,} is beyond a reasonable doubt; {Reinhard Genzel and Andrea Ghez received a share of the 2020 Nobel Prize in Physics for this work.} Observations of stars orbiting Sgr A* have also provided new tests of GR \cite{Hees2017,Gravity2018b,Do2019,Gravity2020}}. For our purposes here, Sgr A* and M87* are important primarily because they are nearby by astronomical standards, enabling particularly detailed studies, e.g., Figure~\ref{fig:EHT_images}. (Although M87* is $\sim 2000$ times further away than Sgr A*, the event horizons of the two BHs subtend similar angles on the sky because of the much larger BH mass in M87*.)

{The most widely used technique to study BHs is seemingly paradoxical:  contrary to their name, BHs are the central engines for some of the brightest and most unusual sources of radiation in the Universe. Astronomers use this radiation to study in detail these fascinating objects.   The radiation is produced by gas (outside the event horizon) that is spiraling into the BH via an accretion disk. Accretion converts gravitational potential energy into heat, which in turn can be converted into radiation and outflows of gas; BHs power the most spectacular observational sources via accretion because they have the strongest gravity of any object in the Universe.   Indeed, the energy liberated by gas falling into supermassive BHs at the centers of galaxies can be so large that it  strongly influences the formation of the host galaxy in which the BH resides {\cite{DiMatteo2005}}.    BHs have thus become a central ingredient in our understanding of how the Universe evolved from the simplicity of the Big Bang to the dizzying variety of stars, planets, and galaxies we see today {\cite{Bruggen2002, Cattaneo2009,Fabian2012}}.} {The birth and growth of supermassive BHs in the first billion years after the Big Bang remains a major puzzle; new observational facilities, such as the James Webb Space Telescope, are likely to shed new light on this longstanding problem.}

{Accretion disks around BHs are a major focus of research in astronomy.   They involve rich physics including a complicated interplay of relativity, gravity, plasma physics, hydrodynamics, and electricity and magnetism.   BH accretion disks, {and their associated mass ejections in the form of relativistic jets,} emit radiation across the entire electromagnetic spectrum, from radio waves to gamma-rays, and are likely sources of high energy neutrinos and cosmic rays.   Much of the radiation we observe originates close to the event horizon, where most of the gravitational potential energy of the inflowing matter is released.  As a result, accretion disks around BHs are also one of our best tools for studying physics in strong gravity, where manifestations of GR are expected to be most prominent.}

{Gas pulled in by a supermassive BH will typically carry substantial angular
momentum {(i.e., a sense of rotation)}, which will prevent the gas from falling directly into the hole. {However,
magnetic field lines embedded in the plasma can transfer angular momentum between fluid parcels. This enables gas to lose angular momentum and to spiral in towards the center via an accretion disk {\cite{Balbus1991}}.} 
A large fraction of the gravitational potential energy
released by accretion is converted to heat (this is inevitable because of the ``friction" associated with angular momentum transfer), and thence to the radiation we
observe (this last step is more tricky than one might think).}

{In a luminous system like a quasar, the conversion of heat to
radiation is straightforward. As gas flows in, at each radius it
self-consistently heats up to just the temperature needed to radiate away the added heat. As the gas approaches the BH, it reaches a temperature
$\sim 10^5$\,K for a supermassive BH, and $\sim 10^7$\,K for a
stellar-mass BH. The former temperature
leads to radiation primarily in the optical and UV bands,
just like we observe in quasars, and the latter gives
X-rays, as in the brightest
stellar-mass BHs.   In this
``standard" disk model {\cite{Shakura1973, Novikov1973}}, the radiation emitted at each radius is roughly of blackbody form, and the total
luminosity of the accretion disk is $\sim 0.1
\dot{M} c^2$, where $\dot{M}$ is the mass accretion rate (grams per
second) into the BH. The net result is that about 10\% of the rest mass energy
of the accreting gas is converted to radiation, which makes such an
accretion disk far more efficient than the best nuclear power sources we
have on Earth ($\sim 0.1$\% conversion for a fission reactor and, theoretically, $\sim 1$\% for a maximally efficient fusion reactor).}

\subsection*{Hot Accretion}

{In astrophysics, an important property associated with an object of
mass $M$ is its Eddington luminosity, $L_{\rm Edd} = 10^{38}(M/M_\odot
)\,{\rm erg\,s^{-1}}$.  This is the limiting luminosity at which the
outward push on a parcel of ionized gas by radiation just
balances the inward pull of gravity. The quasars and X-ray-bright
BHs mentioned earlier have luminosities between a few and 100\% of
$L_{\rm Edd}$.

{Already in the 1970s it was clear that something strange happens with BH accretion when
the luminosity falls below about 1\% of $L_{\rm Edd}$. Stellar-mass
BHs no longer radiate like $10^7$\,K blackbodies, but instead have much hotter gas with
$T \sim 10^9$\,K {\cite{Esin1997}}. Supermassive BHs similarly change in
character. They become unusually dim relative to the mass available for accretion {\cite{Fabian1988}}, and they do not radiate like $10^5$\,K blackbodies {\cite{Ho2008}}. Notably, Sgr A* produces hardly any radiation in any of the expected bands for a standard disk {\cite{goldwurm1994, grindlay1994}}; it is extremely dim, but at the same time appears to have gas at a temperature $T\sim10^{10}$\,K {\cite{Krichbaum1998, Shen2005}}. 

The question is: How can an accretion disk be hugely
hotter than a standard disk and yet have a lower (sometimes far lower) luminosity? The
solution to this problem lay in the identification of a very different mode of accretion called ``hot accretion" {\cite{Yuan2014}}.}

{The first breakthrough came with the recognition that it is possible to have a hot accretion model if the accreting material consists of a two-temperature plasma, a gas in which electrons are significantly cooler than ions (here ``ion" refers to the ionized nucleus of  an atom, in practice, mostly ionized hydrogen, i.e., protons). A two-temperature model of accretion was first developed in a seminal 1976 paper by Shapiro et al {\cite{SLE1976}}.  In this model, since ions are too massive to radiate, they retain most of their viscous heat, reaching temperatures close to the virial temperature, $T\sim10^{12}$\,K near the BH. The pressure associated with this large ion temperature maintains the gas in a tenuous low-density state. The low density in turn ensures that thermal coupling between ions and electrons via Coulomb collisions (the electric force) is weak, thus allowing the plasma to retain its two-temperature character. The electrons meanwhile radiate the energy they receive directly through viscous heat or by Coulomb collisions with the ions; the latter isn't much because of the low density. The low density also makes the gas relatively transparent to its own radiation, and this allows the system to avoid radiating like a blackbody. Overall, the model appears to be just what is needed to explain Sgr A*, M87*, and a host of other low-luminosity BHs. Unfortunately, it has a fatal flaw: it is thermally unstable {\cite{Pringle1976, Piran1978}}, which means that the system cannot survive for any length of time.}

{The second crucial breakthrough came with the recognition that there are in fact two different two-temperature hot accretion solutions, one being the above thermally unstable model, and a second which is stable. The existence of two distinct hot solutions was first recognized by Ichimaru in a forgotten 1977 paper {\cite{Ichimaru1977}}, 
{with related ideas in later work {\cite{Rees1982}}}; it was then rediscovered in the 1990s {\cite{NY1994, NY1995, Abramowicz1995}} and developed in detail. The stable second solution underlies most current models of hot accretion {\cite{Narayan1998,Yuan2014}}. One of its features is that the net radiative luminosity of the accretion disk is less than, often much less than,  $0.1 \dot{M} c^2$. These hot accretion solutions are thus described as being radiatively inefficient.}

{The first object for which a quantitative hot accretion model was developed was Sgr A* {\cite{Narayan1995}}. This early model predicted ion and electron temperatures near the BH of $T_i \sim 10^{12}$\,K, $T_e \sim (1.1-1.3)\times10^{10}$\,K, and a very low radiative efficiency, $L/\dot{M}c^2 \sim 0.0004$. A year later, a similar model was developed for M87* {\cite{Reynolds1996}}, with $T_e \sim 2 \times 10^9$\,K, and radiative efficiency $\sim 0.001$. In the years since, the hot accretion model has been the mainstay of this field, not only for Sgr A* and M87*, but for all low-luminosity supermassive and stellar mass BHs (which are by far the dominant class of accreting BHs in the Universe) {\cite{Ho2008,Yuan2014}}.

{While the hot accretion model is consistent with available observations, we should bear in mind that it is not necessarily correct.  The {model requires} the key assumption of a two-temperature plasma, with outrageously hot ion and electron temperatures, hundreds of times hotter than the center of the Sun in the case of electrons, and yet another factor of hundred hotter for ions. Magnetized plasmas are known to have numerous microscopic instabilities, many with rapid growth rates. It is hard to believe that none of these instabilities is able to tap into the free energy associated with the vastly different ion and electron temperatures and drive the plasma to a single temperature. If even one instability were to succeed in equilibrating the ion and electron temperatures, a key foundation of the hot accretion scenario would collapse, and the entire model would cease to be viable. As we describe in the next section, recent observations have provided new support for the two-temperature hot accretion model and have allayed some of the doubts.}

\subsection*{Inward Bound}

{{Two observational developments have provided a close-up view of the near-horizon environment of Sgr A* and M87*.} The first  is the continued progress of long-baseline interferometry, in the incarnation of the Event Horizon Telescope (EHT) {\cite{EHT_M87*}}.  The EHT is a set of radio telescopes spread across the Earth that can operate as a single instrument by collecting and storing the full electric field information of the radiation as it hits the telescopes.  This technique, interferometry {\cite{Thompson2017}}, allows the resulting array of telescopes to obtain an angular resolution of $\lambda/D$, where $\lambda$ is the wavelength of the radiation and $D$ is the distance between pairs of telescopes.    Fortuitously, for $D$ comparable to the diameter of the Earth and $\lambda \simeq 1$ mm (the shortest radio wavelength where the Earth's atmosphere does not inhibit such observations), the resulting angular resolution is just sufficient to see event horizon scale structure in M87* and Sgr A*.   Equally fortuitously,  M87* and Sgr A* emit copious radiation at mm wavelengths so the sources are conveniently bright at just the wavelength where horizon-scale observations can be made! Finally, if these objects possess the kind of hot accretion flow suggested by models, then the gas would be relatively transparent at $\lambda \simeq 1$ mm, which means that we can look all the way down to the BH horizon without any obscuration by the accreting gas.}

{The second key observational development is the GRAVITY interferometer, in which the European Southern Observatory's four 8m VLT telescopes operate as an interferometer, creating images with 10 times higher angular resolution than any VLT telescope alone could produce {\cite{Eisenhauer2008}}. GRAVITY observes in the infrared at $\lambda = 2\,\mu$m and has made spectacular contributions to our understanding of the Galactic Center, including precise measurements of the BH mass and confirmation of GR via the precession of the orbit of the star S2 around the BH {\cite{Gravity2020}}.   Unlike the EHT, the GRAVITY interferometer cannot directly resolve event-horizon scale structure.  Instead, however, it measures the position of the center of light to an  accuracy comparable to that of the size of the event horizon.}

{Figure \ref{fig:EHT_images} shows the EHT images of M87* and Sgr A*. The general features of these images are remarkably consistent with theoretical predictions made well before the observations:   a deficit of light at small radii in the shadow region of the image {\cite{Falcke2000}}, surrounded by a somewhat asymmetric bright ring of emission.   Two primary constraints need to be satisfied to explain the EHT results. The first is that the radiation from the accreting gas must be produced  within a region at most a factor of few larger in size than the event horizon.  In this case, the combined effects of the gravitational bending of light, which produces the shadow, and Doppler shifts due to the high speed of the gas, which produces the asymmetry in the ring, robustly produce images like those observed. The second constraint is that in order to explain the relatively modest azimuthal asymmetry around the ring, we must be viewing the rotating gas {somewhat} face-on, 
so that most of its orbital motion is not directed towards the observer at Earth.  This is not surprising for M87*, in which there is independent evidence for such a viewing angle {\cite{Walker2018}}. For Sgr A*, however, the implication is that the gas flowing into the BH has a sense of rotation that is somewhat orthogonal to that of the Galaxy as a whole!  Earlier, the GRAVITY interferometer had provided evidence for just such a viewing angle.  A few times a day the BH in our Galactic Center undergoes dramatic ``flares" in which the infrared and X-ray fluxes can increase by a factor of $\sim 10-100$ in less than an hour {\cite{Baganoff2001, Genzel2003, Ghez2004}}.  During one such flare, GRAVITY observed motion of the center of light of the infrared emission {suggestive of} gas orbiting near the BH {\cite{Gravity2018}}.   To cleanly see such orbital motion on the sky requires a {viewing angle closer to face-on than edge-on}, similar to that {favored by} the EHT data.}

{Although the images in Figure \ref{fig:EHT_images} viscerally convey the literally light-warping environment near a BH, the most detailed constraints on theoretical models of accretion physics come, not from the images of total flux, but rather from the polarization of the observed radiation.   The mm and infrared emission observed in Sgr A* and M87* is synchrotron radiation, produced by energetic electrons spiraling around magnetic field lines.   Such radiation is predominantly linearly polarized, with the direction of polarization emitted by the plasma perpendicular to the local magnetic field direction.  Observations of the polarized radiation by the EHT and GRAVITY are so constraining because the polarization directly provides information about the magnetic field direction near the event horizon, and because the polarization direction is also influenced by the plasma through propagation effects that modify the polarization in a plasma density- and temperature-dependent manner.}

\begin{figure}[!ht]
\centering

\includegraphics[width=\linewidth]{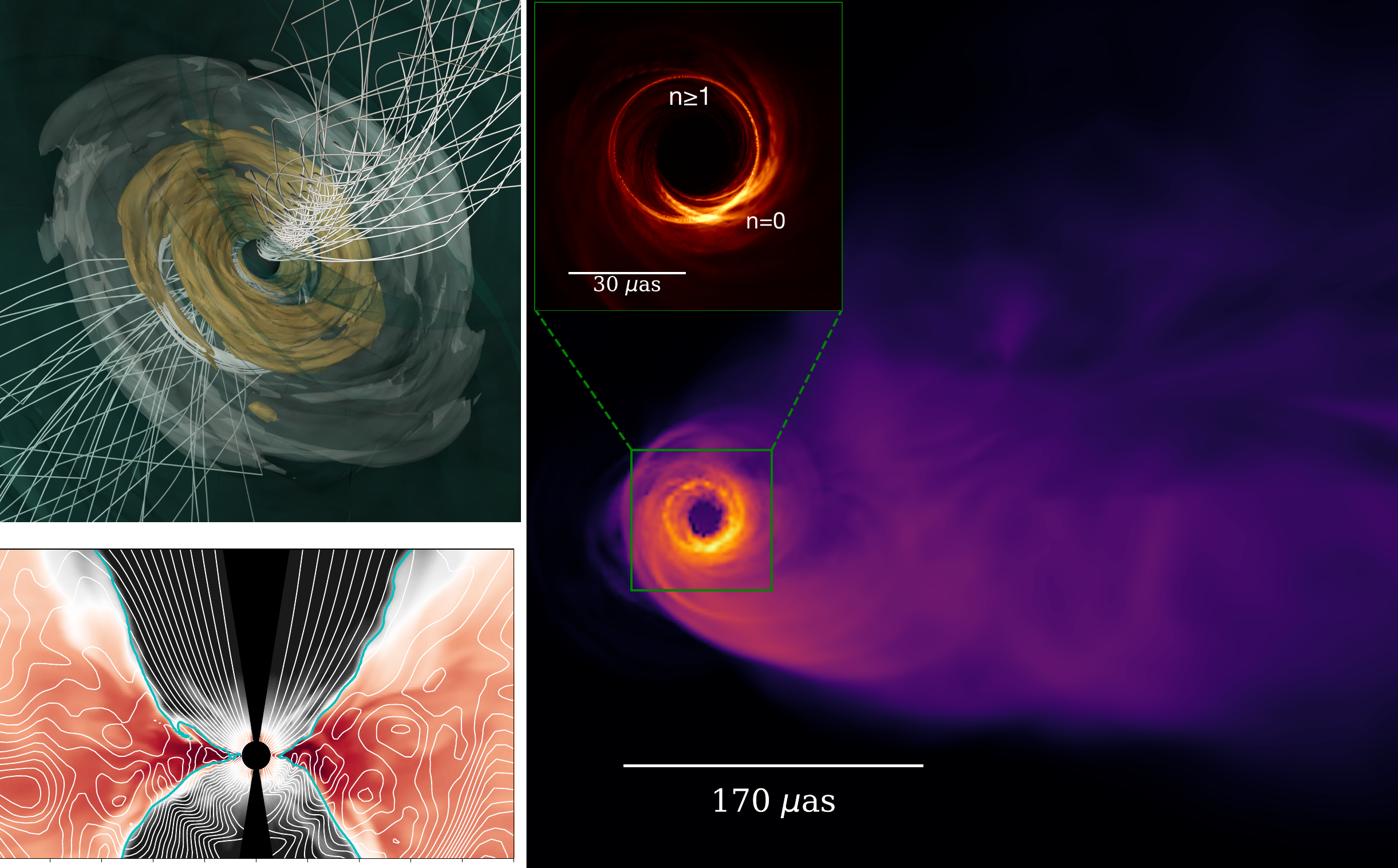} 

\caption{{{\bf Numerical simulation of black hole accretion and the resulting theoretically predicted image}. {\bf Lower Left},  Edge-on view of gas density (red:  high density; black: low density) and magnetic field (white lines) in a GR simulation of gas accreting onto a rotating BH.   {\bf Upper Left},  3D rendering of the magnetic field (white lines) and accretion disk, showing the twisting of magnetic field lines in the jet due to the rotation of the BH. {\bf Right}, Theoretically predicted synchrotron image for M87* using the simulated plasma properties. Inset shows brightness of radiation on a linear scale similar to current EHT data {(with subrings defined in Fig \ref{geodesics} labeled with integer $n$}), while the main figure shows brightness on a logarithmic scale appropriate for future  mm interferometers; the latter can directly reveal the connection between the near-horizon environment and the jet powered by BH rotation.  Figure courtesy of Andrew Chael and George Wong.}}\label{sim}
\end{figure}

{Quantitatively interpreting many features of the EHT and GRAVITY data requires comparing theoretical models of the expected radiation to the observations.   In doing so, the community relies primarily on numerical models of plasma flowing into a BH; see Figure \ref{sim} for an example.   Such models solve the equations for a magnetized fluid in the  space-time of a rotating BH {\cite{Gammie2003,deVilliers2003}}.   The simulations then predict the temperature, magnetic field strength, density of gas, etc. as a function of space and time.   Models of radiation produced by the plasma together with solutions to the transport of photons from the vicinity of the BH to a distant observer are then used to construct synthetic data for comparison to observations (we note that the techniques to make these predictions were developed well before the EHT observations {\cite{Monika2009,Dexter2010}}).   There are some uncertain ingredients in such models, e.g., how the simulations treat the electron vs. proton temperature, what the structure of gas flowing towards the BH at large distances is, etc.  Nonetheless, the theoretically favored set of assumptions prior to the EHT observations have been remarkably successful at explaining most of the observational results.}

{Notably, a key conclusion of the theoretical modeling of EHT data on Sgr A* is that it is very hard to reconcile a single temperature model with the observations; the plasma is very likely two-temperature {\cite{EHT_SgrA*V}}. We do not yet have a direct measurement of the ion temperature, so this conclusion is model-dependent. Nevertheless, the EHT results now put the two-temperature hot accretion model on a firmer footing.

On a more quantitative note, the observed ``brightness temperature" (shown by the color scale in Figure~\ref{fig:EHT_images}) in the EHT images of Sgr A* and M87* correspond to electron temperatures {$T_e > 1.3 \times 10^{10}$\,K in Sgr A* and $T_e > 5\times 10^9$\,K in M87*, close}  to the values predicted by models developed a quarter century ago. {(The measured brightness temperature is a lower limit on the electron temperature; in the case of Sgr A* and M87*, the electron temperature is probably a factor of a few larger than the brightness temperature.)}

{The detailed comparison of theoretical models to observations also constrains the accretion rate $\dot{M}$ onto the BH, because the accretion rate sets the density and magnetic field strength in the gas and thus many properties of the observed radiation.  
For Sgr A*, in particular, the inferred $\dot{M}$ \cite{EHTM87VIII} is
consistent with theoretical inferences made over two decades ago on the basis of the first detections of linear polarization in the mm radiation {\cite{Agol2000, Quataert2000}}. The current EHT-favored models of Sgr A* and M87* have radiative efficiencies of $\sim 10^{-3}$ and $\sim 10^{-2}$, respectively, again surprisingly close to long-ago predictions. Especially in Sgr A*, most of the accretion energy apparently goes through the horizon, rather than emerging as radiation. That is, the hot accretion flow in Sgr A*  is radiatively inefficient, as predicted.}

\subsection*{Testing Gravity with the EHT}

{The region close to a BH and just outside the event horizon is where manifestations of GR are strongest and where gravity deviates the most from Newtonian physics.
Here curvature of spacetime becomes extreme, space is ``dragged" by
the BH's spin, and the quintessential defining feature of a
BH, its event horizon, forms. The EHT, with its ability to
make images of this region, allows us an unprecedented opportunity to
study GR.} {There are few quantitative probes of matter and spacetime under the strong-field conditions near a BH, so observationally diagnosing the near-horizon environment via interferometry is a remarkable opportunity to better understand the  predictions of GR.} In principle, such observations could also reveal hints of the new physics required by the incompatibility of GR and quantum mechanics.

{We begin by discussing the event horizon. If Sgr A* or M87* were to possess a
traditional surface rather than an event horizon, one might expect
some of the mm wavelength radiation emitted by the accretion flow to bounce
off the surface and to produce a visible feature precisely in the dark shadow region of the images in Figure~\ref{fig:EHT_images}. The absence of such a feature in
current EHT images already rules out certain kinds of surfaces, and
the constraints will become tighter in the future. More generally, any
surface would absorb the thermal and mechanical energy dumped on
it by the hot accretion flow and would reradiate this energy. By combining EHT observations with infrared constraints from GRAVITY and other instruments, very
strong limits can be put on such surface radiation, making it increasingly probable that Sgr A* and M87* have event horizons and are thus true BHs {\cite{EHT_SgrA*VI}}.}

{Moving on to spacetime curvature outside the horizon, this is physically parameterized by
the metric, a mathematical construct which describes the geometry of
spacetime. GR predicts a unique form for this metric -- the Kerr
metric \cite{Kerr1963} -- with only two parameters: $M$, the mass of the BH, and $a_*$, a dimensionless number between 0 and 1 which quantifies the spin of the BH. (In
principle, a third parameter, $Q$, the electric charge of the black
hole, is allowed {\cite{Newman1965}}, but astrophysical BHs are unlikely to have
large enough $Q$ to make a difference {\cite{Blandford1977}}.) 
{The prediction of GR that a BH is fully described by just two numbers makes a BH fundamentally different from any other type of astronomical object; e.g., to describe a planet requires specifying its mass, rotation, radius, number of mountains, atmosphere properties, how big the oceans are, etc.}   A demonstration that the spacetime near a BH is perfectly described by the Kerr metric, and that no additional terms are present in the metric, is the holy grail of classical GR. The EHT has now brought us
substantially closer to achieving this goal.}

{At first sight, it would appear that the predicted signatures of
gravity in the observed image of an accreting BH would be
inextricably lost in the complicated and not-fully-understood
details of the accretion process. {Fortunately, this is not the case.  GR predicts that there are unique signatures of strong gravity in the image {\cite{Darwin1959, Bardeen1973, Luminet1979}}.  Gravitational bending of light is so strong around a BH that there is a characteristic radius, the photon orbit, at which even light must travel in a circular orbit around the BH.  For radiation emitted by plasma close to the BH, GR predicts that some of the light in the image ends up in ``subrings" characterized by an integer n:   how many times the light undergoes half of a roughly circular orbit around the BH; see Figure~\ref{geodesics}.  Remarkably, } the sizes and shapes of the subrings are determined entirely by the metric; the only role of accretion is to light up these features for us to make observations. Thus, the sub-rings provide a direct map of the metric.}

{One worry is that the various subrings in the image overlap and the $n>0$ subrings carry only a small fraction of the
total flux. (In fact, the higher-order subrings,
which are the cleanest probes of the metric, have exponentially
suppressed fluxes.) Does this make the entire enterprise impractical?
Fortunately, it does not. In a brilliant insight, Johnson and Lupsasca \cite{Johnson2020}
showed that interferometry separates out the signals from different
subrings and places them in different and easily distinguished regions of data space (specifically, in different interferometric baselines). This allows us the opportunity to study subrings individually despite their
low fluxes.}

\begin{figure}[!ht]
    \centering

\includegraphics[width=\linewidth]{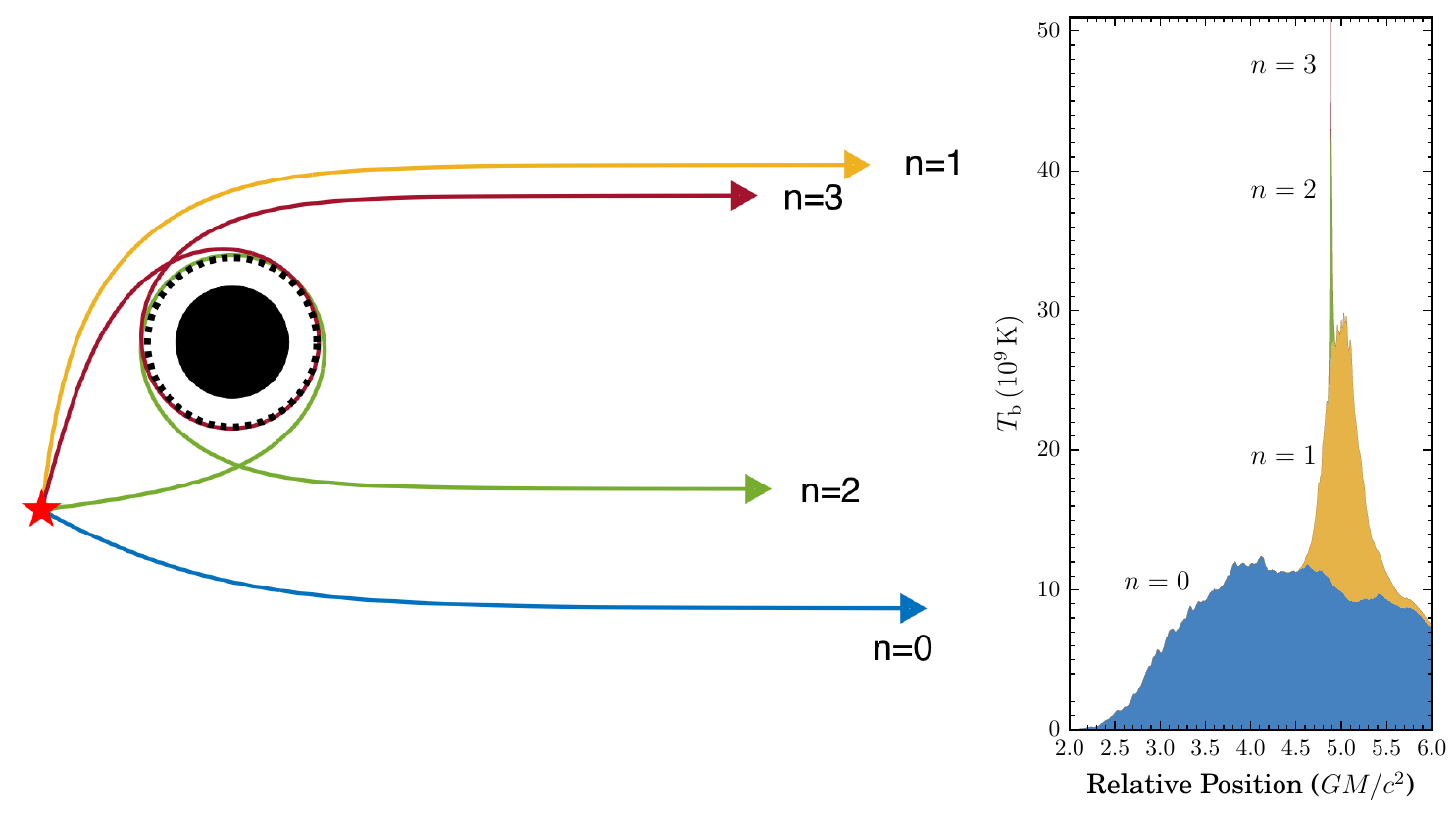} 
    
\caption{{\bf The origin of photon sub-rings. }  {\bf Left},  The path of light rays from a single emitting point traced forward in GR to an observer at far right.  The light trajectories are labeled by the integer number $n$ of half-orbits the photons undergo around the BH. Adding the contribution from all radiating plasma, the observer sees an image consisting of a concentric set of rings; {the $n=0$ rays will produce a relatively broad diffuse ring, while the $n=1$ and higher rings will be progressively sharper (see Fig.~\ref{sim})}. {\bf Right},   Brightness in the image as a function of position (this is a theoretical zoom-in on the brightness near the rings in Fig \ref{fig:EHT_images}); the projected location of the center of the black hole is at an x-axis position of 0 in this Figure, where the flux is very small (corresponding to the dark `shadow' in Fig. 1).  The different subrings in the image, corresponding to the trajectories in the left panel, are labeled.  The observed properties of the $n \ge 2$ subrings are determined primarily by GR, relatively independent of the accretion flow properties.  At the current EHT resolution, what we observe is dominated by the direct $n = 0$ image which has most of the flux, along with some contribution from $n=1$ (see Fig. \ref{sim}.)   Future observations will be able to quantitatively determine the structure of the $n>0$ rings, providing a strong test of GR. Right panel courtesy of George Wong.}\label{geodesics}
\end{figure}

There are a number of proposals for precisely testing the Kerr nature of the BHs in Sgr A* and M87*.    In the currently favored accretion geometries, the shape of the dark shadow at the center of the EHT images depends on both BH mass $M$ and spin $a_*$ so that a higher fidelity image may allow both parameters of the BH to be independently measured, as well as any deviations of the spacetime from the Kerr metric to be searched for {\cite{Broderick2014, psaltis2020, Chael2021, EHT_SgrA*VI}}.  Measuring the shapes of the photon sub-rings is an even more robust approach since those features of the image depend solely on the space-time, not the accretion flow structure {\cite{Johannsen2010,Gralla2020}}.

{The goal of testing the Kerr nature of the BH space-time using  astronomical observations of Sgr A* and M87* is incredibly exciting.   It will undoubtedly require arrays with more telescopes (to see fainter features in the image) and higher angular resolution (to study the narrow photon sub-rings, Figure~\ref{geodesics}).  The latter is particularly challenging and requires going to either higher frequencies (where Earth's atmosphere does not cooperate) or longer baselines.   Longer baselines will ultimately require satellites in space to resolve the $n \ge 2$ photon sub-rings.}

\subsection*{{Energy Extraction from Spinning BHs}}

{In the popular imagination, a BH is an object that eats whatever it can lay its hands on and returns nothing to the external Universe. While largely true, there is an important exception. In a penetratingly simple analysis, Penrose {\cite{Penrose1969}} showed that a spinning BH has free energy associated with its
rotation which can be accessed from outside the hole. He described
an explicit example involving particles in orbit around the black
hole, which showed one way in which this energy might be extracted. The
mechanism depends on frame-dragging, the extraordinary prediction
of GR that space exterior to a spinning object is dragged in the
direction of rotation.}

{While the relevance of Penrose's particle-based toy example to
astrophysics is unclear {\cite{Bardeen1972}}, a related mechanism which involves magnetic fields \cite{Blandford1977} looks very promising. Consider the magnetic field configuration shown in the left panels of Figure~\ref{sim} in which a bundle of magnetic field lines is bunched around the rotation axis of a spinning BH. Because of frame-dragging, the regions of these field lines close to the equatorial plane will be pulled around by the BH, whereas the same lines farther out will feel no such pull. This will induce spiral outgoing waves in the field lines, as in the upper left panel of Figure~\ref{sim}, which will carry energy and angular momentum to large distances {\cite{Semenov2004}}. Correspondingly, the BH will spin down, losing angular momentum and mass energy.} 

{Accreting BHs, especially of the supermassive variety, are
well-known for their relativistic jets: powerful collimated outflows
of radiation and magnetized plasma moving out in twin
oppositely-directed beams at close to the speed of light. The
prototypical example is M87*, which has a famous jet (see Figure~\ref{fig:M87jet}) whose power exceeds the radiative power of the BH's accretion
disk. {The energy carried by such jets far away from the BH and into the surrounding galaxy turns out to be a major source of heating that strongly influences how the surrounding galaxy grows.}} 

{It is tempting to associate relativistic jets with the magnetic
version of the Penrose process. Indeed, there is strong theoretical
support for this proposal since computer simulations of hot accretion onto
spinning BHs show that something similar to the
observed jets forms spontaneously under fairly generic conditions {\cite{Tchekhovskoy2011}}. Moreover, in some cases, the mechanical power in the simulated jet is larger than 100\% of $\dot{M}c^2$, far larger than the $0.1$\% (e.g., Sgr A*) or $1$\% (M87*) that comes out as radiation from a hot accretion disk, or even the $\sim10$\% (typical quasar) from a radiatively efficient standard disk. If the Penrose process is indeed the power source behind the multitude of relativistic jets
found in the Universe, and if we could prove this, it would imply that rotating black
holes all over the Universe are returning a part of their energy to the Universe,
and that they are doing so via frame-dragging, a central prediction of GR.}

{Can we prove specifically that the jet in M87* gets its power by bleeding spin energy from the BH? This would be
a spectacular verification of frame-dragging and the Penrose
process.  

\begin{figure}[!ht]%
\centering

\includegraphics[width=\linewidth]{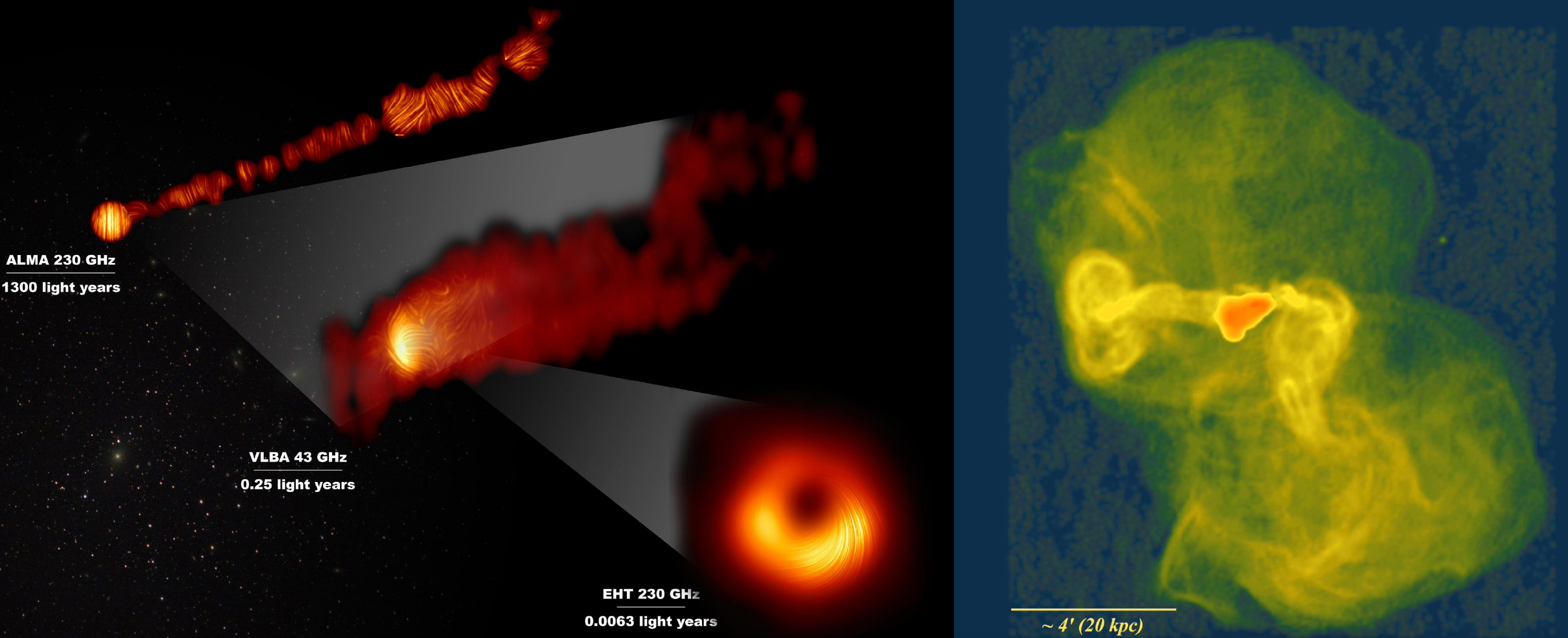} 

\caption{{\bf Radio/mm images of the relativistic jet in M87*}. {\bf Left}, The three images {\cite{EHT_M87VII, Goddi2021, Kravchenko2020}} go from a scale of {$0.006$ light year} in the EHT image at the bottom, which is similar to the size of the BH horizon, to a scale of 1000~light years at the top, comparable in  size to the core of the M87 galaxy. {\bf Right}, Large scale image {\cite{Owen2000}} which covers an area $100,000$~light years across (kpc = 3,260 light years), showing that the jet stirs up a volume that includes the whole galaxy and beyond.  
Credit left panel: EHT Collaboration; ALMA (ESO/NAOJ/NRAO), Goddi et al.; VLBA (NRAO), Kravchenko et al.; J. C. Algaba, I. Martí-Vidal; right panel: F.N. Owen, J.A. Eliek and N.E. Kassim, National Radio Astronomy Observatory, Associated Universities, Inc. \copyright AAS. Reproduced with permission. DOI: 10.1086/317151}
\label{fig:M87jet}
\end{figure}

{The combination of EHT images, polarization, and, better yet, movies based on time-resolved images, appears to be a promising tool to do this.  For example, frame dragging likely imprints signatures on the shapes of polarized images, which may be observable in the future {\cite{Ricarte2022}}.} {The harder step would be to show that frame-dragging is the primary
reason for the presence of the jet and that the energy actually flows out of the spinning BH.}

Theoretical models indicate that the outflowing power in a jet powered by the magnetic version of the Penrose process depends on both the BH spin $a_*$ and the amount of magnetic flux threading the BH horizon $\Phi_{BH}$, with jet power $\propto a_*^2 \Phi_{BH}^2$ {\cite{Blandford1977, Tchekhovskoy2011}}. So how powerful can a jet become?
There is a limit to how rapidly a BH can spin; the limit corresponds to $a_*=1$, which loosely speaking is when the horizon rotates at the speed of light. But is there a limit also to how big $\Phi_{\rm BH}$ can grow? If not, Nature could make jets arbitrarily powerful by increasing $\Phi_{\rm BH}$ without bound. It turns out that there is in fact a limit on $\Phi_{\rm BH}$. If we try to squeeze more and more field lines close to the event horizon, the magnetic pressure builds up and pushes new lines away. The limiting magnetic flux is reached when the outward push of the magnetic field just matches the inward push of the accreting gas. A system in this limiting state is called a magnetically arrested disk (MAD). All other things being equal, a MAD system will have the most powerful jet. This maximum jet power is determined purely by the BH spin $a_*$ and the mass accretion rate $\dot{M}$ {\cite{Tchekhovskoy2011}}. 

{The models that successfully explain EHT observations of M87*, especially the polarization data (Figure~\ref{fig:EHT_images}), require M87* to be in the MAD state {\cite{EHTM87VIII}}. This is consistent with the presence of a powerful jet in this system. GRAVITY- and EHT-based models indicate that Sgr A* also may be in the MAD state. Is it possible that most hot accretion flows in the Universe are MAD? This is certainly plausible since, once a system has reached the MAD limit, the field lines are pinned to the BH by the inflowing gas. The only way to eliminate the field is by bringing in magnetic flux of the opposite sign and canceling out what is already there. The new flux should have the precise magnitude to avoid under- or over-shooting, which seems unrealistic.}

{Curiously, Sgr A* does not show any {obvious} sign of a jet. If this system is in the MAD state, as models suggest, then a plausible explanation for the lack of a {powerful} jet is that the BH spins very slowly. This hypothesis will be tested once the BH spin is measured.  {Alternatively, it could be that a jet is actually present {\cite{FalckeMarkoff2000,Yusef-Zadeh2006,Brinkerink2015}} but is stifled by the surrounding environment, which prevents the jet from traveling to large distances from the BH where it could be more readily observed.}}

\section*{Looking Forward}

{In reflecting on the diverse science enabled by the initial EHT and GRAVITY results for Sgr A* and M87*, it is inspiring to realize that this is but a taste of what is to come.  We are at the beginning of what will be a many decades journey to ever more precise observations of the environment around BH event horizons. 

{In the future improved infrared interferometery {and spectroscopy} {with bigger telescopes} will detect stars even closer to the BH in Sgr A* and will better characterize the near-horizon environment via {measurements of orbits} and observations of flares.} {If we are lucky, we may discover a radio pulsar in orbit around Sgr A*; precision timing of this pulsar would be a game-changer {\cite{Psaltis2016}}.}
New generations of mm interferometers, both on the ground and  in space, will reveal far more details of the near horizon environment.  This will enable stringent tests of GR using the photon sub-rings.  A second key advance will be directly observing emission somewhat farther from the BH, elucidating the connections between the accretion inflow, the (likely) spinning BH, and the outflowing jet, all of which are critical for understanding the broader role of BHs in the formation of structure in the Universe.} 

{In parallel to these advances in our understanding of BHs using light, observations of gravitational waves from merging BHs (a topic beyond the scope of this Perspective) will continue to improve dramatically.  The timing of radio pulsars in our Galaxy is approaching the needed sensitivity to detect the gravitational wave background produced by mergers of $\sim 10^9 M_\odot$ BHs {\cite{PTA2020}}.  Next generation ground-based detectors will be able to observe stellar-mass BH mergers across the cosmos {\cite{Kalogera2021}};  detailed studies of the brightest of these events will test the uniqueness of the Kerr BH solution.  In the 2030s, the space-based LISA gravitational wave observatory will be launched and will detect mergers of $\sim 10^{5-7} M_\odot$ binary BHs, as well as stellar-mass objects spiraling in towards a massive BH
{\cite{LISA2017}}.}

\subsection*{{Acknowledgments}}
We are grateful to Andrew Chael, Daniel Palumbo, and George Wong for their help making Figures 1-3.


\newpage

\section*{{Data availability}}

{EHT data for Figure 1 are available at https://github.com/eventhorizontelescope.}

}

\section*{{Code availability}}

{Figure 1 and the right panel of Figure 2 were created using the ehtim (eht-imaging) software package which is available at https://github.com/achael/eht-imaging {\cite{Chael2018}}.}

}

\section*{Competing interests} 

The authors declare no competing interests.

\section*{Author contributions}

R.N. and E.Q. contributed equally to all aspects of the paper.

\end{document}